— LIGHTWEIGHT USABLE CRYPTOGRAPHY —
A USABILITY EVALUATION OF THE ASCON 1.2 FAMILY

Arne Padmos

We present a usability study of the Ascon 1.2 family of cryptographic algorithms. As far as we know, this is the first published experimental evaluation aimed at a cryptographic design (i.e. not a specific API) with the purpose of informing which aspects to standardise. While the results show the general difficulty of choosing and applying cryptographic algorithms, there are some more specific insights. These include the possibility of confusing multiple variants, the relevance of small interfaces, and the desire for higher-level wrapper functions (e.g. for protocols). Overall, many questions are still open, including how usability could be integrated into the design and evaluation of cryptographic algorithms. Our main takeaway is that lightweight usable cryptography is an open research area that deserves greater focus. For the review of NISTIR 7977, the standardisation process of Ascon as a FIPS, and when exploring potential future SPs, the key criterion of usability should be based on realistic user testing and on triangulation from other methods.

INTRODUCTION

In 1999, the seminal usable-security user study into PGP 5.0 by Whitten & Tygar showed that usability is an important barrier when it comes to real-world security. Between the '90s and now, the usable security field has built up a rich and varied history. The kinds of topics studied have increased in diversity, while the field has also become more mainstream over the years.

In the area of cryptography, besides studies related to email encryption and secure messaging, there has been a handful of studies in the area of cryptographic APIs. These studies show that outside of a focus on end users, developers have also been considered. However, as far as we are aware, no studies have yet been published that approach cryptographic designs from a usable security perspective with the goal of supporting standardisation processes. This is peculiar as NISTIR 7977, published back in 2016, notes the importance of usability as one of nine principles that guide the standards development process of NIST.

NISTIR 7977 states that 'cryptographic standards and guidelines should be chosen to minimise the demands on users and implementers as well as the adverse consequences of human mistakes and equipment failures'. As such, we ran a user study on the Ascon 1.2 family of cryptographic algorithms to identify potential pitfalls.

## METHODOLOGY

The population of the study was students following a second-year undergraduate course on information security. Over 80% of students registered for the course participated in the study. The original intent was for students to work in pairs, but some students ended up working alone while others worked in larger groups. Around 55 students participated, leading to around 20 to 25 unique submissions.

The task was to create a Python implementation of a protocol for securing the communication between a glucose monitor and an insulin pump. To ensure that each participant had a clear mental model of the problem domain as well as the relevant security threats to address, the study started with a group discussion in each of the three study runs. Every discussion consisted of creating a data-flow diagram (DFD), identifying crown jewels, key assumptions, and trust boundaries, after which the STRIDE mnemonic was used to elicit what could go wrong. For each run, three DFDs were sketched: data sent from the sensor to the pump, with and without acknowledgement, and data fetched from the sensor by the pump.

On the basis of the data-flow diagrams and a list of threats, it was up to the participants to create an implementation. They were linked to https://pypi.org/p/ascon and asked to take three steps of stepwise refinement: defining relevant message flows, writing pseudocode, and coding in Python. Participants were encouraged to move on to the next step every 10 minutes. They were also encouraged to refer to the threats that were previously identified. After around 30 minutes, results were emailed and participants were provided with individual feedback the same week.

## RESULTS

None of the participants managed to implement a complete and correct solution. Based on exploratory initial qualitative analysis of the solutions, several recurring issues can be observed. The Ascon AEAD functionality provided by the PyAscon implementation was often called with zero, one, or just a few parameters instead of the full parameters specified in the documentation at https://pypi.org/p/ascon. Also, it was common for solutions to include a wrapper function around exiting functions in the PyAscon library. This wrapper function commonly took only the message to be transmitted as input. Relevant cryptographic parameters such as the key, nonce, associated data, etc. were often hard-coded or missing entirely (appearing out of thin air when the PyAscon functions were called). Aspects such as error handling and key diversification were missing from all solutions.

Besides the above general patterns, there were also more specific highlights from individual submissions. One solution called the Ascon-128 variant for encryption and the Ascon-128a variant for decryption. Another called the Ascon-80pq variant with a 128-bit key without good reason (this worked due to a bug in the reference implementation, making a call specifying the Ascon-80pq variant behave like the Ascon-128 variant). One solution disregarded the PyPI documentation and instead invented their own object-oriented API (consisting of creation of an Ascon object and related 'encrypt', 'send', 'receive', and 'decrypt' methods). Some solutions showed confusion between cryptographic message authentication codes and the terminology of MAC addresses.

Interestingly, while the group discussion included a consideration of threats to confidentiality, some solutions described only countermeasures for authentication and not encryption of messages. A single solution included a detailed overview of relevant threats from the group discussion, but this didn't appear to provide much help in arriving at a working solution. Only one solution included a description of a protocol based on counters and the keeping of state. While the general protocol flow appeared to provide a relevant basis for a correct solution, the submission did not include refinement into Python code using the PyAscon library.

## DISCUSSION

The solutions to the exercise illustrate the difficulties that developers face in the proper selection and application of cryptographic algorithms. Confusion around the term 'MAC' indicates that prior knowledge can interfere with cryptographic concepts, suggesting that those who name cryptographic functions should think about potential confusion with similar terms from other fields. Another aspect relates to the number, type, and ordering of parameters to functions. Automatic generation of parameters like nonces – especially when these are 160 bits long – as well as defining standard serialisation formats such as nonce+ciphertext+tag may reduce the need for a multitude of parameters.

The results indicate that there are many open questions regarding the interfaces provided by cryptographic building blocks as well as how these are instantiated in (the API of) reference implementations. For example, is the parameters order of some function signatures more natural than others? Do programming paradigms have any influence on the likelihood of making different kinds of mistakes? Can compatible user-friendly wrappers be created? Especially this last question seems to be a challenge that is very hard to address in a simple and clean manner unless usability is considered early on and at a fundamental level. Developers that have clarity as to the general protocol concepts involved and a clear idea of the threats to be addressed should not have to reinvent the wheel.

## LIMITATIONS

The participants to this study were students, which might not be representative of those using Ascon out in the field. Those designing cryptographic protocols might be more experienced – e.g. when it comes to products like Signal – but we would claim that the participants of this user study are an overestimate of the capability and effort that most IoT developers will put into the selection and application of cryptographic algorithms.

The task included one library applied to a single use case. For the latter concern, this study could be repeated with a different use case, e.g. communication between smart lights and a hub. As to just one implementation being tested, we note that this reference implementation has been developed by one of the Ascon designers, which prevented the researcher carrying out the user study from developing an implementation in line with prior expectations. On the other hand, expanding on insights from the usability field, more usable interfaces could be designed prior to evaluation. Based on audience feedback at the 2023 Permutation-based Crypto workshop, how and when to integrate usability considerations into development life-cycles of cryptographic algorithms as well as how designs should be evaluated from a usability perspective during a competition are still open questions.

Note that this user study involved interpretation of artefacts. Having participants 'think aloud' while they complete these exercises might provide more details as to the thoughts behind the outcomes. However, besides such an improvement to laboratory experiments, it appears valuable to explore complementary usability methods. The field of ethnography has provided a rich source of inspiration for various user-experience research techniques. Similar approaches might be useful when it comes to gaining insights into lightweight usable cryptography.

## CLOSING THOUGHTS

Usability is dependent on the user, task, and context. Standardisation aims to find a common denominator. Whereas the former has a grounding in approaches such ethnography, the latter may be more engineering oriented. During standardisation processes, an important pitfall to consider is evaluating (simplicity of) designs in too restricted a context, which could lead to not considering the added complexity and overhead within broader systems. Relatedly, before standardising features that may turn out to be foot guns or foot cannons, evaluation should be performed in a realistic context of use for a given application domain in order to properly weigh risks and benefits.

Similar considerations apply when it comes to the relevant features to consider. Historically speaking, protocol security has been a mess. Mistakes were even found years after the publication of protocols by renowned authors (some modes such as OCB2 suffered the same fate, with published proofs later turning out to be faulty). In light of this history, it seem appropriate to consider where and how Ascon will be used. At first glance, two-party half-duplex record protocols with ratcheting are an interesting test case for evaluating whether the proposed Ascon modes and parameters provide extensibility for a common setting.